\documentclass[preprint]{aastex62}%twocolumn
\usepackage{amssymb,amsmath,graphicx,natbib,hyperref}
\usepackage{color}
\usepackage{enumerate}
\usepackage{multirow}

\usepackage{ulem}
\usepackage[encapsulated]{CJK}
\usepackage[utf8x]{inputenc}
\DeclareGraphicsExtensions{.pdf,.PDF,.png,.PNG,.jpg,.JPG,.jpeg,.JPEG}
\newcommand{\hi}{{\rm H}{\textsc i}~}
\newcommand{\him}{{\rm H}{\textsc i}}
\graphicspath{}
\defcitealias{Kruk2018}{K18} 

\shorttitle{correlation between \hi and bars}
\shortauthors{Zhou et al.}

\begin{document}
\begin{CJK*}{UTF8}{gbsn}

\title{On the Correlation Between Atomic Gas and Bars in Galaxies}
\author{Zhimin Zhou(周志民)}
\affiliation{Key Laboratory of Optical Astronomy, National Astronomical Observatories, Chinese Academy of Sciences, Beijing, 100012, China}
\correspondingauthor{Zhimin Zhou}
\email{zmzhou@nao.cas.cn}

\author{Jun Ma}
\affiliation{Key Laboratory of Optical Astronomy, National Astronomical Observatories, Chinese Academy of Sciences, Beijing, 100012, China}
\affiliation{College of Astronomy and Space Sciences, University of Chinese Academy of Sciences, Beijing 100049, China}

\author{Hong Wu}
\affiliation{Key Laboratory of Optical Astronomy, National Astronomical Observatories, Chinese Academy of Sciences, Beijing, 100012, China}
\affiliation{College of Astronomy and Space Sciences, University of Chinese Academy of Sciences, Beijing 100049, China}

%\author{Suijian Xue}
%\affiliation{National Astronomical Observatories, Chinese Academy of Sciences, Beijing, 100012, China}

%%%%%%%%%%%%%%%%%%%%%%%%%%%%%%%%%%%%%%%%%%%%%%%%%%%%%%%%%%%%%%%%%
%                            ABSTRACT                           %
%%%%%%%%%%%%%%%%%%%%%%%%%%%%%%%%%%%%%%%%%%%%%%%%%%%%%%%%%%%%%%%%%
\begin{abstract}
	We analyze the correlation between properties of large-scale bars and atomic gas content of galaxies to explore the role of \hi gas on bar evolution in galaxies. We show that the absolute bar size depends strongly on total stellar mass of galaxies and does not change significantly with \hi gas fraction at fixed stellar mass. Furthermore, the physical size of the bar is small and nearly constant in high \hi gas fraction and low-mass galaxies, and becomes larger with increasing galactic stellar mass in low gas fraction galaxies. When the stellar masses are fixed, the relative bar length normalized to the disk shows a decrease with increasing \hi gas fraction due to the larger disks in gas-richer galaxies. 
	We measure the gas deficiency of the samples and find that the gas-rich galaxies have longer and stronger bars compared with the \hi gas-deficient galaxies at fixed stellar mass, especially for the massive ones. % This \textcolor{red}{likely} indicates that bars can grow steadily in the disks with a small amount of gas. while they are inhibited in a significant amount of \hi gas.
	When splitting the samples into star forming and quiescent subgroups, the star forming galaxies tend to have longer bars than the quiescent ones at fixed stellar mass and gas deficiency. 
	In addition, our results suggest two different types of bar properties, in which the bars in gas-rich galaxies grow longer but retain similar axial ratio over time, while they grow longer and fatter over time in gas-poor galaxies.

\end{abstract}

%%%%%%%%%%%%%%%%%%%%%%%%%%%%%%%%%%%%%%%%%%%%%%%%%%%%%%%%%%%%%%%%%
%                        INTRODUCTION                           %
%%%%%%%%%%%%%%%%%%%%%%%%%%%%%%%%%%%%%%%%%%%%%%%%%%%%%%%%%%%%%%%%%
\section{Introduction}
\label{sec:intro}
%bars introduction and properties
%the role of bars on galaxy evolution
%the effect of gas on bars
	Galactic bars are elongated structures that extend from the center of galaxies are a ubiquitous and important component of disk galaxies. Observations have indicated that over 60\% of disk galaxies in the local universe contain bars, including both strong and weak bars\hypersetup{citecolor=blue}\citep{Menendez2007, Nair2010, Lee2019}, and bars are still present in about 10\% of the galaxies up to z $\sim$ 1 \citep{Sheth2008, Melvin2014, Simmons2014}.
%	\sout{Beside that, The observed bar fraction may vary due to the classification method, wavelength dependence, selection effects, and physical properties of the host galaxies \citep{Zou2014, Li2017}.}

	Bars are the important internal drivers for secular evolution of galaxies \citep[see][for a review]{Kormendy2004}. They can effectively redistribute the gas and stars within galaxies by large-scale streaming motions \citep{Regan1999, Sheth2002, Athanassoula2003, Sheth2005}, flatten the chemical distribution of the disk \citep{Martin1994, Gadotti2001, Sanchez2014, Fraser2019, Zurita2021}, enhance star formation activity in the central region, and contribute the formation of (pseudo)bulges \citep{Sersic1965, Sersic1967, Hawarden1986, Ho1997, Sheth2005, Ellison2011, Lin2017}.
	%In addition, the inflow of gas induced by bars might reinforce the fueling of the central supermassive black holes and trigger active galactic nuclei (AGN) activity \citep{Oh2012, Galloway2015, Alonso2018, Kim2020}, although no clear correlation between them has been observed so far \citep{Lee2012, Cisternas2013}.	
%	However, bars are not always found to enhance the central star formation or  concentrations of cold gas \citep{Fisher2013, Zhou2015}, they are also postulated to suppress the global or galaxy scale star formation in their host galaxies \citep{Donohoe2019, Lin2020, Newnham2020, Wang2020}. 

	Although bars are ubiquitous in the local universe, their formation and evolution history remains mysterious, a large open questions about their origin, growth, and destruction are still poorly understood. Typically, bars tend to form either through inherent instability of galactic disks \citep{Athanassoula2013, Kwak2017, Zana2018}, through external perturbations such as galaxy$-$galaxy mergers, or through tidal forces from neighboring structures \citep{Berentzen2004, Lokas2016, Yoon2019, Cavanagh2020}.

%	Once a bar has formed, it can grow and evolve while it drives the secular evolution of its host galaxy \citep{Bournaud2002}. However, the factors relevant for the origin and evolution of bars are many and not easy to be disentangled, such as dark matter halos, gas component, baryonic physics and the impact of external processes \citep{Athanassoula2013a, Zhouzb2020}. 

	In addition, the formation and evolution of bars may depend on many factors such as dark matter halos, gas content, and internal and external physical processes of galaxies \citep[e.g.,][]{Sheth2008, Athanassoula2013a, Zhouzb2020}.
	 Multiwavelength observations have shown that bar fraction and structural parameters are correlate with galaxy mass, morphology, color, and gas content \citep[e.g.,][]{Masters2011, Kim2014, Melvin2014, Cervantes2015, Diaz2016}, but the results are still controversial.
	Some studies have shown that bars are more frequently found in massive, gas-poor, and red disk galaxies \citep{Erwin2005, Cheung2013, Gavazzi2015}, while other work reports that bars are tend to exist in less massive, gas-rich, and blue spirals \citep{Barazza2008, Erwin2018}, or have a bimodal distribution on galactic properties \citep{Nair2010, Masters2011, Diaz2016}. Thus, detailed analysis of the correlation between bars and gas content will shed light on the possible mechanisms that could affect the formation and evolution of bars.

	It is well known that gas has a fundamental effect on the evolution of disk galaxies. Gas content, in particular neutral hydrogen (\him) gas, can be accreted onto the disk from the intergalactic medium \citep{Kere2005, Sancisi2008, Nelson2013, Zhou2018}, and then the gas will be driven inflow into galactic central region by bars because of their nonaxisymmetry gravitational potential \citep[e.g.,][]{Athanassoula1992}. Meanwhile, bars evolve during the exchange of angular momentum and can be reinforced, weaken, or be destroyed in the influence of the gas content \citep[e.g.,][]{Kraljic2012, Athanassoula2003, Athanassoula2013}.

%, which is the raw material that forms molecular clouds and then stars, plays a crucial role in galaxy formation and evolution \citep[e.g.,][]{Zhang2009, Zhou2018, Guo2020}. However, its effect on the evolution of bars remains unclear. 
	Many observational works have investigated the dependence of bar presence on HI content \citep[e.g.,][]{Sheth2005, Villa2010, Ellison2011}. \citet{Masters2012} analyzed a sample of local disk galaxies with both bar classifications and \hi measurements and found that the bar fraction is significantly lower among gas-rich disk galaxies than in gas-poor ones, even at fixed color or stellar mass. Similarly, \citet{Davoust2004} showed that barred galaxies have lower \hi mass fractions than the unbarred galaxies. 

	Theoretical works have also demonstrated that the stellar bars form later and are weaker in the presence of large amounts of gas than in gas-poor galaxies \citep{Berentzen2007, Athanassoula2013, Rosas2020}, and they even do not form at all for gas fractions that are higher than 0.3 \citep{Lokas2020}. In addition, the growth of the central mass concentration due to the inflow of gas can cause bars to dissolve \citep{Algorry2017}, and a galaxy may have several bar episodes in its life, including bar formation and renewal \citep{Bournaud2002}. All these results suggest that gas is inhibited in the presence of a strong bar \citep{Combes2008} or it inhibits bar formation \citep{Villa2010}. 

However, it is still difficult to establish an unassailable link between \hi gas and bars, since cold gas is pervasive in galaxy halos and bars are located in the inner region of galaxies. Furthermore, most of the recent observational studies focused on the bar fraction but not the bar properties. Investigating the specific properties of bars will contribute to a deeper understanding of their evolution. In this paper, we aim to use data on bar physical parameters measured by \citet{Kruk2018} and \hi information from the Arecibo Legacy Fast Arecibo L-band Feed Array (ALFALFA) blind \hi survey to study the correlation between \hi gas content and galactic bars. We investigate how the bar properties depend on the gas content of the galaxy and analyze the role of \hi gas on the bar evolution.

This paper is organized as follows. Section \ref{sec:data} introduces the data and galaxy samples. The analysis and results are presented in Section \ref{sec:analysis}. We discuss our results and compare them with previous works in Section \ref{sec:discuss}, and then we provide a brief summary in Section \ref{sec:summary}.

Throughout this paper, we adopt a standard $\Lambda$CDM cosmology with $\it {H}_{\rm 0} {\rm= 70\ km\ s^{-1}\ Mpc^{-1}}$, $\rm {\Omega_M = 0.3}$, and $\rm{\Omega_{\Lambda} = 0.7}$.

%%%%%%%%%%%%%%%%%%%%%%%%%%%%%%%%%%%%%%%%%%%%%%%%%%%%%%%%%%%%%%%%%
%                        Data and Sample                     %
%%%%%%%%%%%%%%%%%%%%%%%%%%%%%%%%%%%%%%%%%%%%%%%%%%%%%%%%%%%%%%%%%
\section{Sample and Data}
\label{sec:data}

%%%%%%%%%%Figure %%%%%%%%%
\begin{figure*}
	\centering
	\includegraphics[width=0.9\hsize]{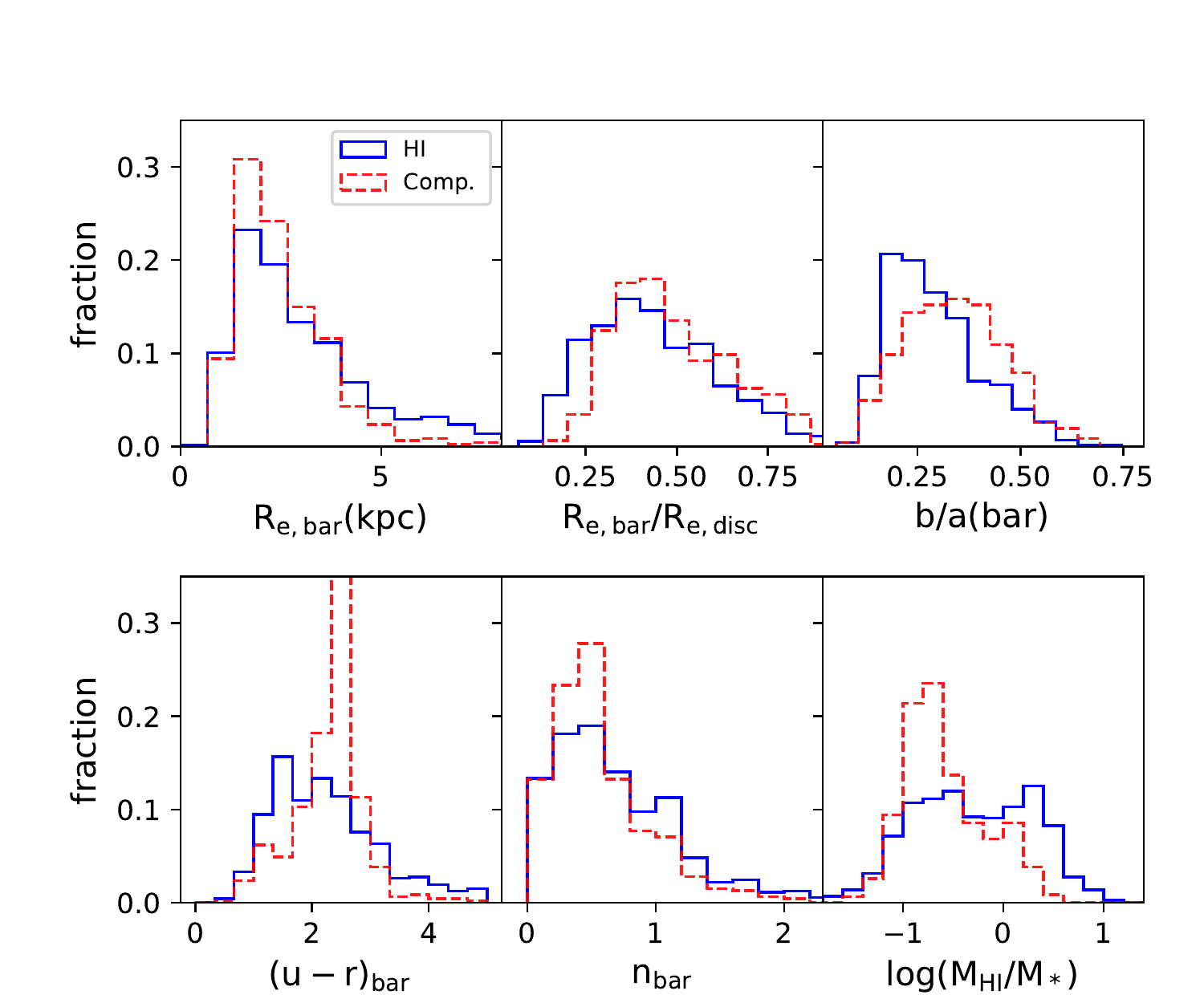}
	\caption{Normalized distributions of the bar structural parameters and \hi gas fraction for our sample galaxies. The blue solid histograms represent the \him-detected barred galaxies, and the red dashed ones indicate the comparison sample without \hi detection. The gas fraction $\rm log(M_{\him}/M_*)$ represents estimated limiting values for the comparison sample.
\label{hist_bar}}
\end{figure*}

\begin{figure}[h]
	\centering
	\includegraphics[width=0.6\columnwidth]{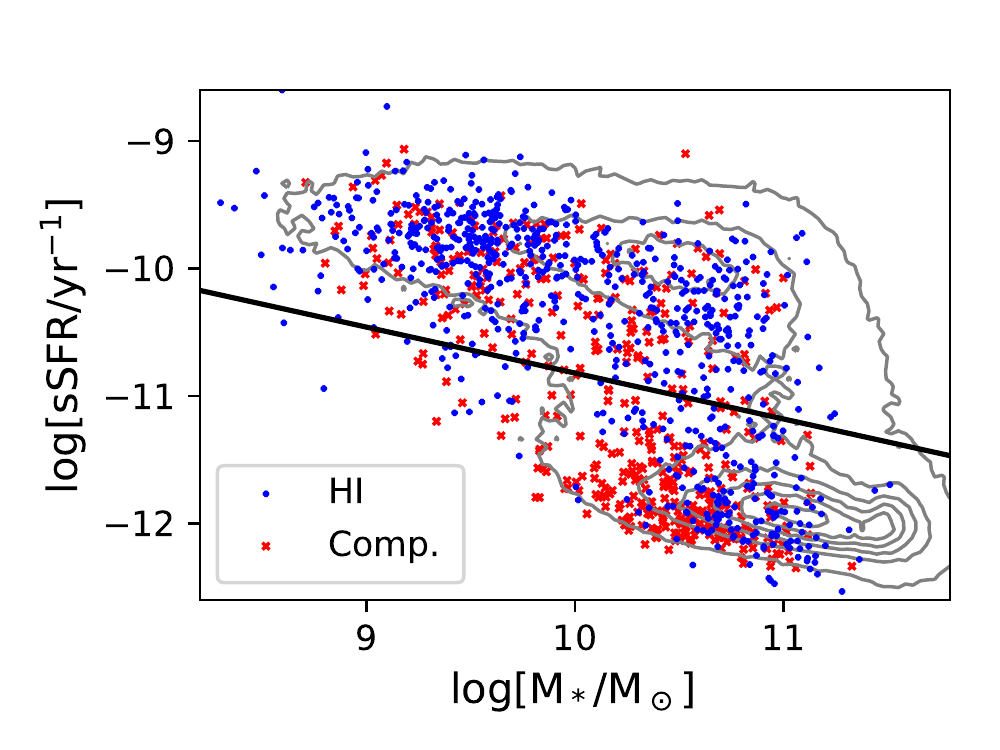}
	\caption{The sSFR$-$mass diagram for our samples. The \him-detected sample and its comparison sample without \hi detection are marked with blue circles and red crosses, respectively. The contours present the distribution of galaxies from the MPA-JHU DR7 database. The solid line, defined by \citet{Woo2013}, shows the division between the blue cloud or star forming galaxies (above the line) and the red sequence or passive galaxies.
	\label{dist_sfr_ms}}
\end{figure}

%%%%%%%%%%Figure %%%%%%%%%%%%

\subsection{ALFALFA}
The ALFALFA survey used the Arecibo 305 m radio observatory in Puerto Rico to map over $\sim$ 7000 deg$^2$ of the sky at high Galactic latitude, yielding a database of 21 cm \hi line sources with 5.3 galaxies deg$^{-2}$. It can detect \hi\ masses as low as 10$^6$ $\rm M_{\odot}$, and cover the recessional velocity range up to 18,000 km s$^{-1}$ with a spectral resolution of 10 km s$^{-1}$. The more detailed information about ALFALFA can be found in \citet{Giovanelli2005}, \citet{Haynes2011} and \citet{Haynes2018}.

\citet{Haynes2018} present the completed catalog of $\sim$ 31,500 extragalactic \hi line sources detected by the ALFALFA survey out to z $<$ 0.06. The catalog includes both detections with high signal-to-noise ratio (S/N$>$6.5) and the sources that have low S/N ($<6.5$) but coincide with likely optical counterparts at the same position and redshift. The \hi\ mass $\rm M_{\hi}$ in the catalog is computed via the standard formula
\begin{equation}
\rm M_{\hi} = 2.356 \times 10^5 D^2_{H} S_{21},
\end{equation}
where $\rm D_H$ is the distance to the galaxy, in units of Mpc, and $\rm S_{21}$ is the integrated \hi\ line flux density of the source, in units of Jy km s$^{-1}$.

\subsection{The bar decomposition catalog}
The morphological parameters used in this work are obtained from \citet[][hereafter \citetalias{Kruk2018}]{Kruk2018}, who performed multiwavelength 2D photometric decompositions for a sample of barred galaxies. 
It was originally selected from Galaxy Zoo 2 \citep{Willett2013}, which is a citizen scientist project with morphological classification of galaxies. Galaxy Zoo 2 measures morphological features such as bars, bulges, disks, and spiral arms through visual inspection of images drawn from the Sloan Digital Sky Survey (SDSS). Based on the morphological classifications from Galaxy Zoo 2, \citetalias{Kruk2018} selected galaxies with bar likelihood $\it p_{\rm bar}$ $\ge$ 0.5, similar to the classic strong bar classification, and excluded merging galaxies or those with high inclinations $i \ge 60\arcdeg$. Their final sample includes $\sim$ 3500 barred galaxies in the redshift range of 0.005 $<$ z $<$ 0.06 and with Petrosian magnitude of $m_r$ $<$ 17.

The structural decompositions of \citetalias{Kruk2018} are performed by GALFITM, which can fit a wavelength-dependent model with multiple components to images in different bands \citep{Bamford2011, Vika2013}. For a given galaxy, two (disk+bar) or three (disk+bar+bulge) components are fitted on the images of five bands ($\it ugriz$) from SDSS DR10 \citep{Ahn2014}.

In addition to the bar structural parameters and \hi information, we also include  galaxy properties from the MPA-JHU DR7 database \citep{Kauffmann2003, Brinchmann2004, Tremonti2004} and retrieve aperture- and extinction-corrected stellar masses $\rm M_*$ and star formation rates (SFRs) of galaxies.

\subsection{Sample and galaxy properties}
There are $\sim$ 2000 \citetalias{Kruk2018} barred galaxies located in the survey region of ALFALFA. We cross-match the barred sample of \citet{Kruk2018} with the ALFALFA catalog with a maximum searching radius of 4\arcsec. The typical FWHM of SDSS images is $\sim$ 1.\arcsec5, corresponding to 1.6 kpc at the redshift $\it{z} \rm{= 0.05}$. Thus, we rejected the galaxies, whose effective radius of the bar major axis ($\rm R_{e,bar}$) is less than 1.\arcsec5 in images. \rm This yields 727 barred galaxies with \hi detections, which compose the \him-detected sample of this study. There are also 1191 \citetalias{Kruk2018} barred galaxies in the ALFALFA survey region but without \hi detections.

For the \him-detected galaxies, we directly obtain their \hi masses from the ALFALFA catalog and then measure their \hi deficiency as
\begin{equation}
\rm def_{\him} = \langle log(M_{\him}/M_*)\rangle-log(M_{\him}/M_*),
	\label{hidef}
\end{equation}
where $\rm \langle log(M_{\him}/M_*) \rangle$ is the expected gas fraction for a given stellar mass and given by the trend of the HI gas fraction with stellar mass in \citet{Masters2012}:
\begin{equation}
\rm \langle log(M_{\him}/M_*) \rangle = -0.31 - 0.86[log(M_*/M_{\odot}) - 10.2].
	\label{meanfhi}
\end{equation}

For the galaxies without being \hi detected, we calculate their upper-limiting \hi flux density $\rm S_{21}$, upper-limiting \hi mass, and lower-limiting \hi deficiency $\rm def_{\him,lim}$ following the method in \citet{Masters2012} and equations \ref{hidef} and \ref{meanfhi}. Next, we define the galaxies with $\rm def_{\him,lim} >0$ as the comparison \him-poor sample. Because there are contaminations from the human-generated radio frequency interference in the \hi detection of ALFALFA, some sources above the detecting limits are missed in the ALFALFA survey \citep{Martin2010, Haynes2018}. Thus, the comparison sample only includes the galaxies within the redshift range of 0.02 $\le$ z $\le$ 0.05. Finally, we identify 467 sources of \him-poor barred galaxies. In addition, the comparison sample should be similar to or more \him-poor than the \him-detected galaxies with $\rm def_{\him}>0$.

Figure \ref{hist_bar} presents the distribution of the bar properties in the sample, including the absolute length, scaled length, axis ratio, color, and S\'ersic indices. The colors have been corrected for Galactic dust extinction and $\it k$-corrected. We did not consider the projection correction due to galaxy inclination for structural parameters. Because nearly face-on galaxies are chosen in our sample, its effects should be minimized. The \hi gas fraction of the galaxies is also shown in Figure \ref{hist_bar}, with the upper-limiting \hi gas mass fraction for the comparison sample overlapped. Our samples span a wide range of physical properties, and the \him-poor galaxies show similar distributions of the bar parameters.

Figure \ref{dist_sfr_ms} shows the distribution of our galaxies in the specific SFR (sSFR)$-$mass diagram. There is a clear bimodality in the stellar mass owing to our galaxies spanning the classic blue cloud and red sequence. More than a half of the \him-detected galaxies are located in the blue cloud, and the comparison galaxies are mainly in the red sequence.

%%%%%%%%%%%%%%%%%%%%%%%%%%%%%%%%%%%%%%%%%%%%%%%%%%%%%%%%%%%%%%%%%
%                   ANALYSIS AND RESULTS                        %
%%%%%%%%%%%%%%%%%%%%%%%%%%%%%%%%%%%%%%%%%%%%%%%%%%%%%%%%%%%%%%%%%
\section{Analysis and Results}
\label{sec:analysis}
In this work, we focus on investigating the correlation between bar properties and \hi gas, and we are also concerned with the effect of bars on star formation of galaxies, which might be able to provide some clues to understanding the evolution process of bars and bar-driven secular evolution. 

%%%%%%%%%%Figure %%%%%%%%%
\begin{figure*}
	\centering
	\includegraphics[width=\hsize]{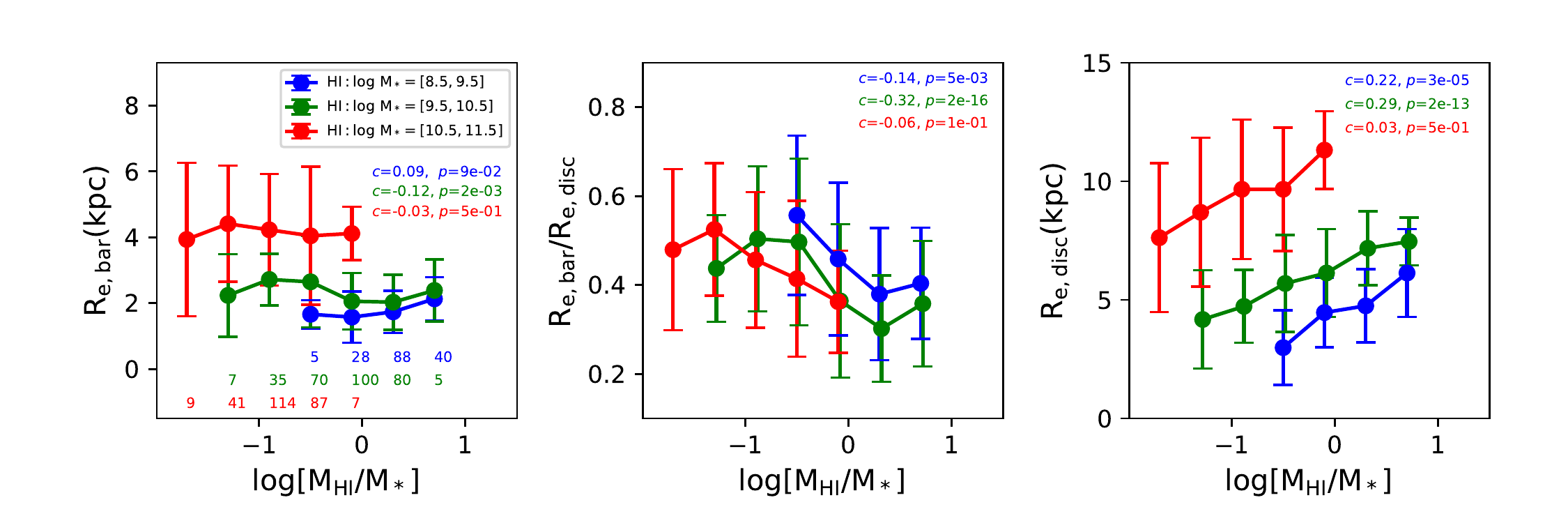}
	\caption{Distribution of bar and disk size as a function of the gas fraction for the \hi-detected sample. The effective radius of each structure is used as the absolute physical size, and their ratio is used as the relative size of the bar. The sample is divided into three groups: high mass ($\rm 10^{10.5}-10^{11.5}$), intermediate mass ($\rm 10^{9.5}-10^{10.5}$) and low mass ($\rm 10^{8.5}-10^{9.5} M_{\odot}$). In the panel, the three groups are marked with different colors; large filled circles and error bars are the mean value and its error for each mass and gas fraction bin. The galaxy number of each bin is denoted in the bottom of the left panel. The Kendall's $\tau $ correlation coefficient ($\it c$) and the corresponding $\it p$-value ($\it p$) are shown at the top of each panel.
	\label{bl_fhi_Mstar}}
\end{figure*}
%%%%%%%%%%Figure %%%%%%%%%
%%%%%%%%%%Figure %%%%%%%%%
\begin{figure*}
	\centering
	\includegraphics[width=\hsize]{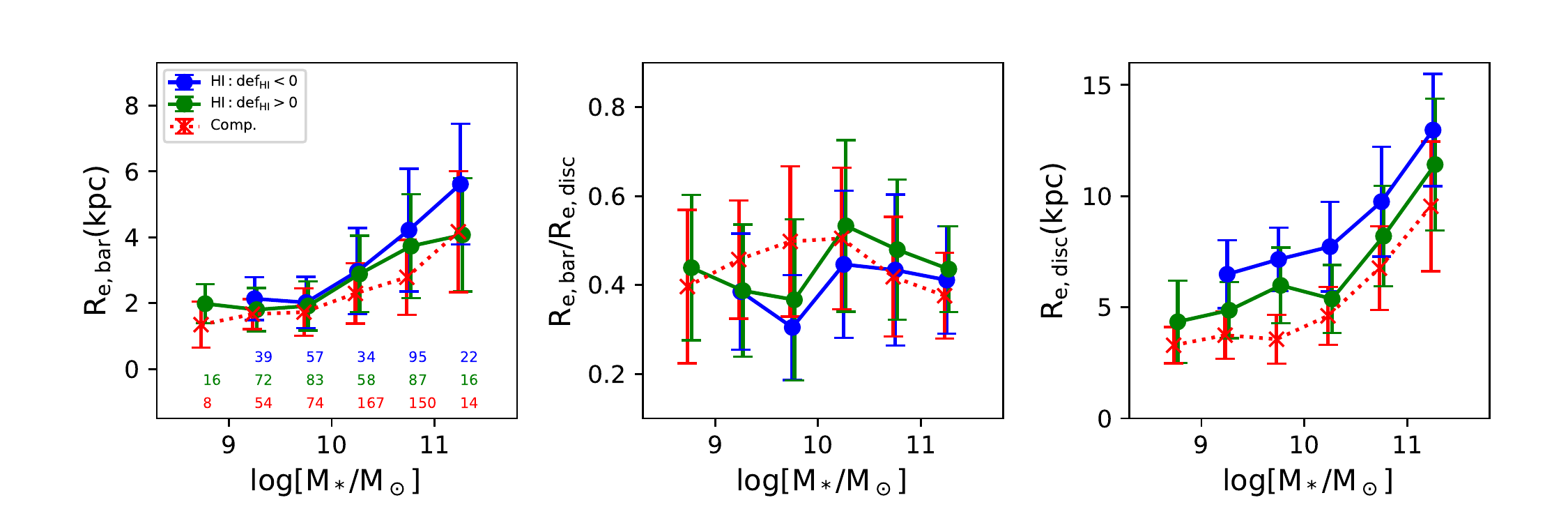}
	\caption{The distribution of bar and disk size as a function of stellar mass. Similar to as Figure \ref{bl_fhi_Mstar}, except that the \hi-detected sample is divided into two groups: galaxies with $\rm def_{\him}<0$ and $\rm def_{\him}>0$, along with the comparison sample ($\rm def_{\him,lim}>0$).
	\label{bl_Mstar_df}}
\end{figure*}
%%%%%%%%%%Figure %%%%%%%%%
%%%%%%%%%%Figure %%%%%%%%%
\begin{figure*}
	\centering
	\includegraphics[width=0.8\hsize]{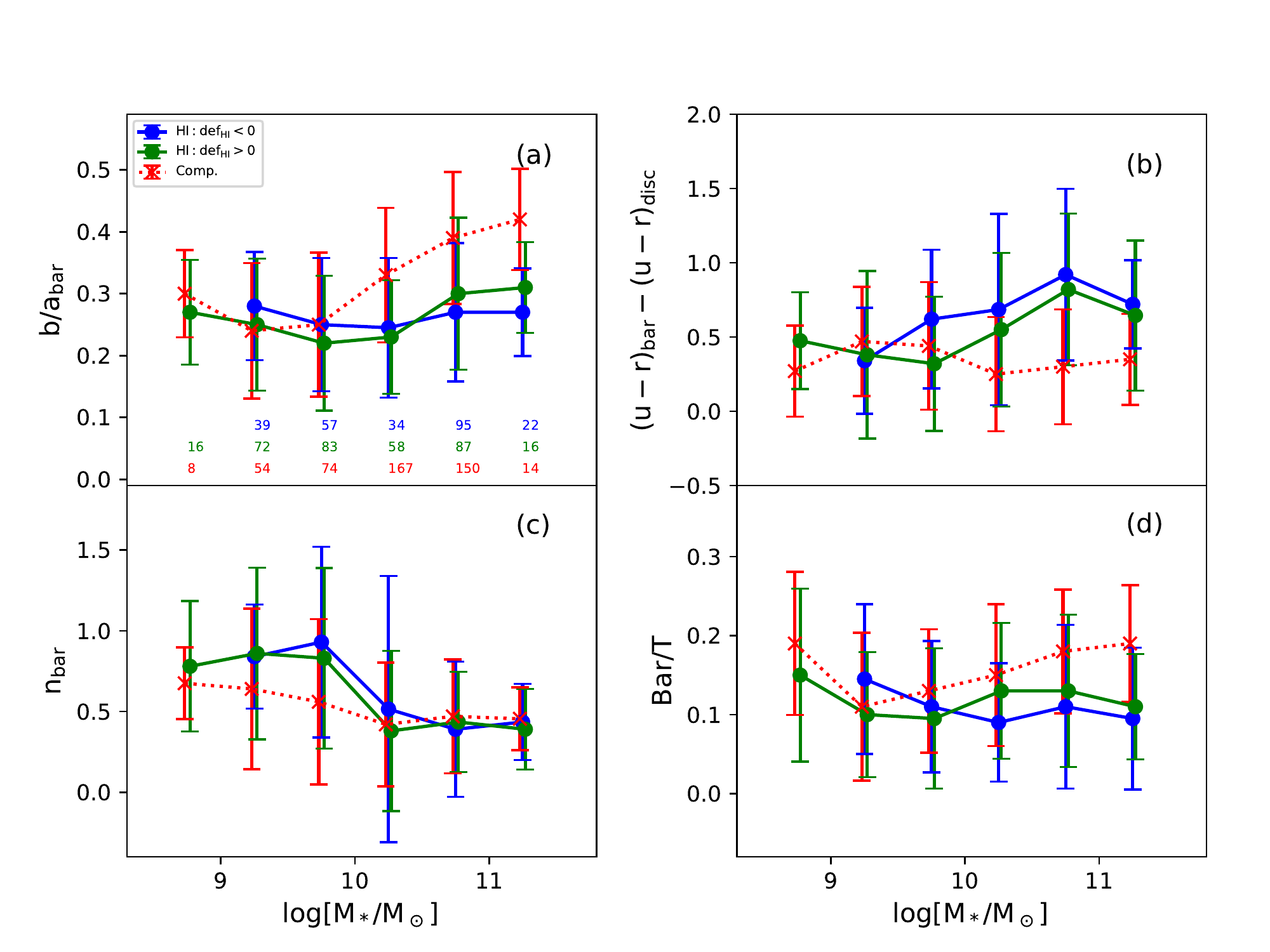}
	\caption{Distribution of physical parameters of bars as a function of the stellar mass. The bar axis ratio $\rm b/a_{bar}$ (panel (a)),  color difference between bar and disk $\rm (g-r)_{bar}-(g-r)_{disk}$ (panel (b)), S\'ersic indices $\rm n_{bar}$ (panel (c)), and bar-to-total luminosity ratio $\rm Bar/T$ (panel (d)) are presented. Same galaxy bins and symbols in Figure \ref{bl_Mstar_df}.
	\label{bp_Mstar_df}}
\end{figure*}
%%%%%%%%%%Figure %%%%%%%%%

\subsection{Bar length with \hi gas}

In Figure \ref{bl_fhi_Mstar}, we show how the absolute and relative bar lengths could be related to the \hi gas fraction $\rm f_{\him} =log[M_{\him}/M_*]$. We use the effective radius of bar $\rm R_{e,bar}$ to trace its absolute length and the bar-to-disk effective radius ratio as its relative lengths. The sample is divided into the high-, intermediate- and low-mass subsamples with the stellar mass in the range of $\rm 10^{10.5}-10^{11.5}$, $\rm 10^{9.5}-10^{10.5}$, $\rm 10^{8.5}-10^{9.5} M_{\odot}$, respectively. For each subsample, we further divide it into different gas fraction bins. The subsamples are marked with different colors in Figure \ref{bl_fhi_Mstar}, along with the galaxy number of each bin denoted at the bottom of the first panel.

%%%%Additionally, the best-fit power-law parameters for each 2pCF are given in Table 3.

The effective radius of the bar shows a wide diversity in the range of 2-6 kpc for the galaxies with lower \hi gas fraction and is within a relatively narrow range around $\sim$ 2 kpc for those with higher gas fraction. The stellar mass bin presents little or weak change in the absolute bar length across all gas fractions; the Kendall's $\tau $ correlation yields a correlation coefficient close to 0.0. In addition, when compared with the lower-mass galaxies, the galaxies with higher stellar masses have longer absolute bar size in the low \hi gas fraction ($\rm log(M_{\him}/M_*)<-0.3$), and have similar bar length in the high \hi gas fraction ($\rm log(M_{\him}/M_*)>-0.3$). This may indicate that the absolute bar length is mainly driven by the galactic stellar mass in gas-poor galaxies and remains constant in gas-rich and low-mass galaxies.

In Figure \ref{bl_fhi_Mstar}, we also see a declining relative bar length toward higher \hi gas fraction, even in the same stellar mass bins. This is not surprising since \hi is widespread in galaxy disks and gas-rich galaxies tend to have large disks \citep{Giovanelli1988, Wang2011}. As we see in the right panel of Figure \ref{bl_fhi_Mstar}, the size of the galaxy disk has a tight positive correlation with both \hi gas fraction and stellar mass, i.e., larger disks tend to be more massive or have higher gas fraction.

In order to further explore the role of \hi gas on the bar growth, we use $\rm def_{\him}=0$ as the dividing line between the \him-rich and \him-poor subsamples and recheck the bar size as a function of galactic stellar mass in Figure \ref{bl_Mstar_df}. This figure confirms the trend of bar and disk sizes with stellar mass and \hi fraction shown in Figure \ref{bl_fhi_Mstar}. However, it also shows that the absolute bar length is slightly larger in gas-rich galaxies ($\rm def_{\him}<0$) than in gas-poor galaxies ($\rm def_{\him}>0$ or $\rm def_{\him,lim}>0$) with the same stellar mass, especially for high-mass bins. This indicates that the massive galaxies likely tend to have longer bars if they contain more gas content than others.

%%%%%%%%%%Figure %%%%%%%%%
\begin{figure*}
	\centering
	\includegraphics[width=\hsize]{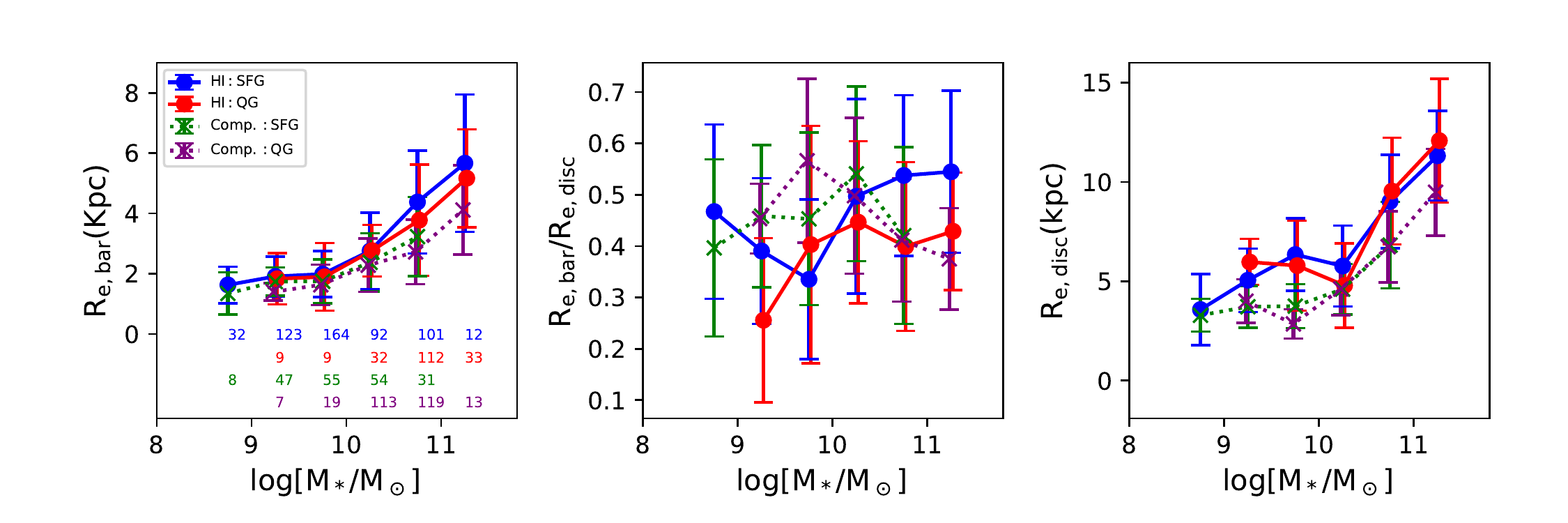}
	\caption{Distribution of the absolute and relative bar lengths as a function of the stellar mass, similar to Figure \ref{bl_Mstar_df}, except that the \him-detected and comparison samples are split into two subgroups by the star formation properties as in Figure \ref{dist_sfr_ms}. Star forming galaxies are marked as $\it H{\small I}: SFG$ (blue circle) for the main sample and $\it Comp.: SFG$ (green cross) for the control sample; the quiescent galaxies are marked as $\it H{\small I}: QG$ (red circle) for the main sample and $\it Comp.: QG$ (purple cross) for the control sample.
	\label{bl_Ms_SF}}
\end{figure*}
%%%%%%%%%%Figure %%%%%%%%%

%%%%%%%%%%Figure %%%%%%%%%
\begin{figure*}
	\centering
	\includegraphics[width=0.8\hsize]{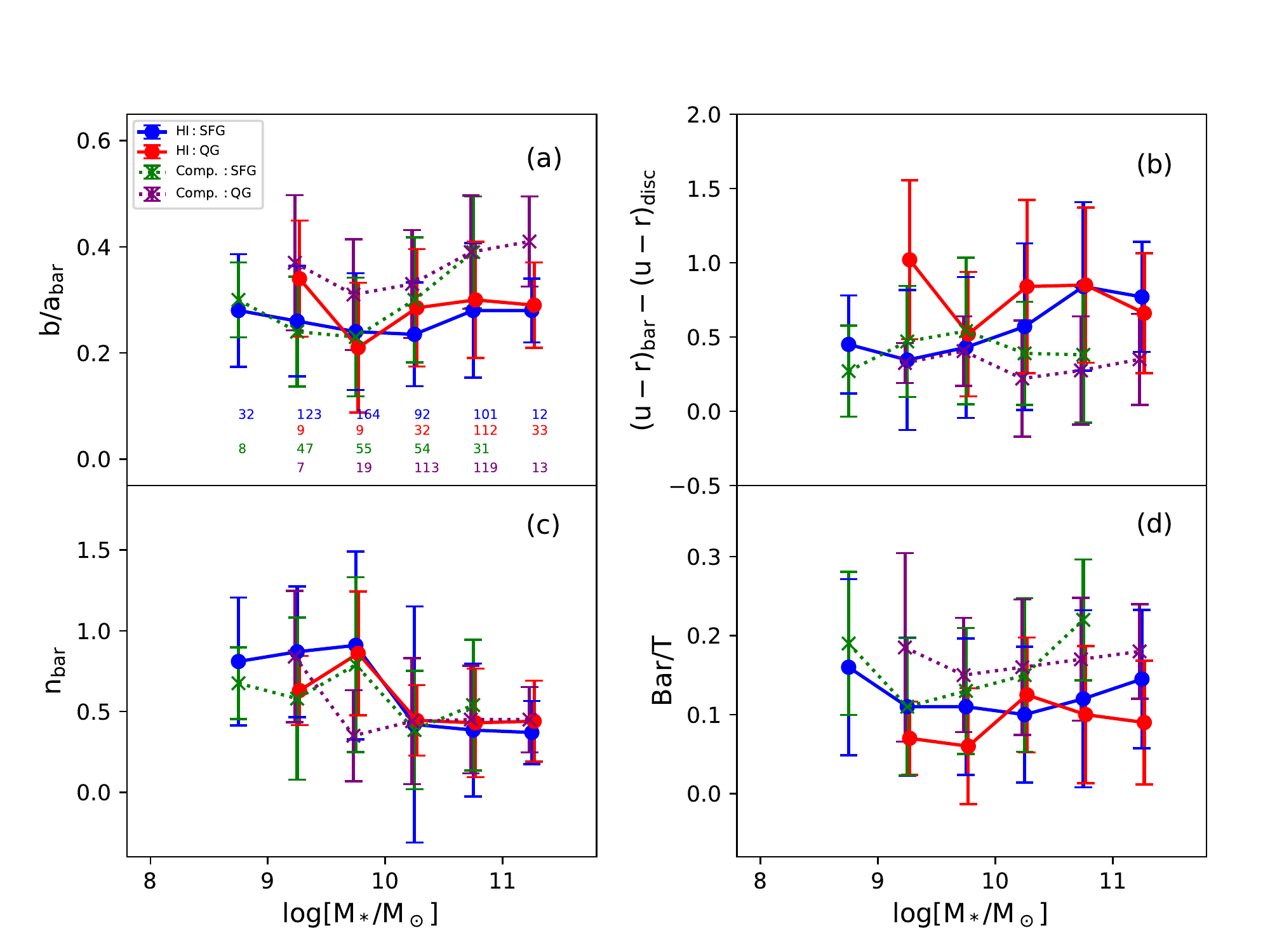}
	\caption{Distribution of physical parameters of bars as a function of the stellar mass, similar to Figure \ref{bp_Mstar_df}, except that the samples are split as Figure \ref{bl_Ms_SF}.
	\label{bp_Ms_SF}}
\end{figure*}
%%%%%%%%%%Figure %%%%%%%%%

\subsection{Other bar properties with \hi gas}
We also explore the physical parameters of bars, including the bar axis ratio $\rm b/a_{bar}$, S\'ersic indices $\rm n_{bar}$, bar-to-total luminosity ratio $\rm Bar/T$ and color difference between bar and disk $\rm (g-r)_{bar}-(g-r)_{disk}$. Figure \ref{bp_Mstar_df} shows the dependence of these parameters on stellar mass and \hi gas across three \hi deficiency bins: $\rm def_{\him}<0$ (blue) and $\rm def_{\him}>0$ (green) for the \him-detected sample, as well as the comparison sample ($\rm def_{\him,lim}>0$, red).

In the four panels of Figure \ref{bp_Mstar_df}, our galaxies show different behaviors in high and low stellar mass bins, respectively. In the high-mass bins, bars in gas-rich galaxies have smaller axis ratio, have lower luminosity ratio, and are redder compared to the disks than those in gas-poor galaxies, while the low-mass bins show little change in these properties across all $\rm def_{\him}$ subgroups. Interestingly, contrary to these parameters, the S\'ersic indices $\rm n_{bar}$ show a wide distribution in low-mass galaxies, with the trend that bars with higher $\rm def_{\him}$ (i.e., gas-poor) have smaller $\rm n_{bar}$ at fixed stellar mass, while $\rm n_{bar}$ remains relatively constant around $\sim$ 1\ in the high-mass bins regardless of being gas-rich or gas-poor. These results likely indicate the different roles of gas and stellar mass in the growth of bars.

\subsection{Possible correlations with star formation}

Besides \hi gas content, the bar itself also evolves along with the star formation in galaxies, enhanced or suppressed. Thus, we illustrate the correlation between star formation and bar properties in Figure \ref{bl_Ms_SF} and \ref{bp_Ms_SF}. For this analysis, we use the division line in Figure \ref{dist_sfr_ms} to split the \him-detected galaxies and comparison gas-poor galaxies into two subgroups, respectively. The sources above the division line are classified as star forming or blue cloud galaxies (marked as $\it H{\small I}: SFG$ for the \him-detected sample, $\it Comp.: SFG$ for the comparison sample), the others are quiescent or red sequence galaxies (marked as $\it H{\small I}: QG$ for the \him-detected sample, and $\it Comp.: QG$ for the comparison sample, respectively).

%Figure \ref{bl_Ms_SF}a
In the left panel of Figure \ref{bl_Ms_SF}, we find that for the massive galaxies ($\rm log M_* >10^{10} M_{\odot}$), both star forming subgroups tend to have larger $\rm R_{e,bar}$ than the quiescent ones. In addition, the star forming and quiescent subgroups of the \him-detected sample both have longer bars compared to their corresponding subgroups of the comparison sample (Figure \ref{bl_Ms_SF}). The former suggests that the stellar bar has positive correlation with the star forming process in galaxies.
%bars are able to grow accompanying the star forming process or star formation could be enhanced by bars. 
The latter is consistent with the result of Figure \ref{bl_Mstar_df} that bars are longer in galaxies with lower gas deficiency at fixed stellar mass. On the other hand, however, these trends nearly disappear for the low-mass galaxies  ($\rm log M_* <10^{10} M_{\odot}$). Furthermore, we also find that the disk sizes at fixed stellar mass show little differences between both subgroups from the same sample, indicating a tighter link between disk and gas than between bar and gas. 

Figure \ref{bp_Ms_SF} presents the variation of $\rm b/a_{bar}$, $\rm n_{bar}$, $\rm Bar/T$, and $\rm (g-r)_{bar}-(g-r)_{disk}$ on different star formation properties. It is worth noting that the bar axis ratio shows different behaviors at high- and low-mass bins. At low stellar mass, bar axis ratios are similar to each other for the subgroups with the same star formation classifications, i.e., $\rm b/a_{bar}$ in \him-detected star forming galaxies is 0.2$-$0.3, comparable with that in the star forming galaxies from the comparison sample, and both of them are smaller than those in the quiescent galaxies. However, at the high-mass bins, their differences almost disappear between the star forming and quiescent subgroups with the same \hi deficiency, and $\rm b/a_{bar}$ are smaller in the \him-detected sample compared in the gas-poor sample. Except for the bar axis ratio, the other parameters such as \rm $n_{bar}$, $\rm Bar/T$ and $\rm (g-r)_{bar}-(g-r)_{disk}$ do not show significant dependence on the star formation property of galaxies but present different distributions in the low- and high-mass galaxies as shown in Figure \ref{bp_Mstar_df}. These results likely imply that multiple potential factors drive their variations.

%%%%%%%%%%%%%%%%%%%%%%%%%%%%%%%%%%%%%%%%%%%%%%%%%%%%%%%%%%%%%%%%%
%                        DISCUSSION                             %
%%%%%%%%%%%%%%%%%%%%%%%%%%%%%%%%%%%%%%%%%%%%%%%%%%%%%%%%%%%%%%%%%
\section{Discussion}
\label{sec:discuss}
% bar in gas: suppression or promotion
% bar bimodal
% bar and star formation w/o gas

% bar size as a Function of \hi gas
We have investigated the correlation between bar properties and \hi gas, stellar mass, and star formation of galaxies. The parameters of bars show different dependence on the galaxy properties. More specifically, the absolute bar length remains short and constant ($\sim$ 2 kpc) in gas-rich and low-mass galaxies and becomes longer (2-6 kpc) with increasing galactic stellar mass in gas-poor galaxies. When the stellar masses are fixed, the relative bar length shows a decrease with increasing gas fraction $f_{\him}$ because that gas-richer galaxies tend to have larger disks. This is consistent with the observational studies on bar fraction. For example, \citet{Masters2012} found an anticorrelation between the bar fraction and the \hi gas richness with SDSS and ALFALFA data, especially for strong or long bars. 

\subsection{How Do Bars Evolve under the Effect of \hi}

Theoretical studies of \citet{Athanassoula2013} have predicted that the relative gas fraction plays an important role in the formation and evolution of the bar, and large-scale bars form much later in gas-rich than in gas-poor disks. In addition, bars easily form and develop in massive and rotation-dominated disk galaxies \citep[e.g.][]{Athanassoula2002}. Two possible causal links between bar and \hi gas are that bars speed up the gas consumption, or \hi gas prevents the formation and growth of bars \citep[e.g.,][]{Gavazzi2015, Lokas2020}, while it is still debated which is the real one. 

%Given the connection between \hi gas fraction and stellar mass \citep{Catinella2010, Zu2020}, the trends between bar parameters and gas fraction might also be affected by the stellar mass more or less. 

Although bars are prevented or evolve slowly in the presence of significant amounts of disk gas, many simulations have suggested that bar strength and length are most likely to grow steadily in the disks with low gas fraction \citep[e.g.,][]{Rosas2020}. Our results also confirm this scenario. As found in Figure \ref{bl_Mstar_df}, the bars in massive galaxies are longer in gas-rich disks ($\rm def_{\him}<0$) than in gas-deficient disks ($\rm def_{\him}>0$ or $\rm def_{\him,lim}>0$). 
The axial ratio of bars can be used as a rough proxy for the bar strength with lower axial ratios for stronger bars \citep{Abraham2000, Laurikainen2007}. Given this, we also found that the bars at fixed high stellar masses are stronger in gas-rich galaxies than those in gas-deficient galaxies (Figure \ref{bp_Mstar_df}). This indicates that a small amount of \hi gas would likely help the growth of bars, while it might inhibit the bar when the amount of gas is significant.

Furthermore, the trends of the bar parameters have noticeable different properties depending on the galactic stellar mass and gas content (Figure \ref{bl_fhi_Mstar}-\ref{bp_Mstar_df}). In the galaxies with low stellar mass and high gas fraction, the bars show no significant variation in their length, axial ratio, and color, while they can vary widely in the galaxies with high stellar mass and low gas fraction. The contradictory properties of bars have been found in lots of previous studies \citep[e.g.,][]{Li2017, Erwin2018, Lee2020}. \citet{Kim2015} used S\'ersic indices to analyze the light profile of a bar and also found that there are two types of surface brightness profiles for bars, flat bars in massive galaxies and exponential bars in less massive galaxies, as shown in Figure \ref{bp_Mstar_df}(c). 
This suggests that the effects of gas likely go a long way toward accounting for the different properties of the bars between in low- and high-mass ones \citep[e.g.,][]{Lee2019}.

%\subsection{star formation via bar}
Although we have not yet addressed the possible causes behind the mechanisms of bar evolution, we try to find some clues from the star formation of galaxies. The effect of the bar on star formation has been addressed by a host of studies \citep[e.g.,][]{Sheth2005, Ellison2011, Buta2015}. They suggested that the nonaxisymmetric potential of  bars can drive gas inflow and trigger vigorous star formation in the center of galaxies, and the bar grows in this process. As expected, we found that the star forming galaxies tend to have longer bars than the quiescent ones (Figure \ref{bl_Ms_SF}). Our further examination also confirms that there is no significant difference in the disk size of the star forming and quiescent galaxies at a given stellar mass and gas deficiency. This result suggests that long or strong bars could enhance the star formation in galaxies, and the bar-driven star formation mainly takes place in the central region of galaxies \citep{Oh2012, Cheung2013}.

However, the variation of bar axial ratio presents significant differences in the star forming and quiescent galaxies with different stellar mass (Figure \ref{bp_Ms_SF}(a)). At low stellar mass, the bar axial ratios are comparable in gas-rich and gas-poor cases with the same star formation classifications. At high stellar mass, the ratios are similar for the galaxies with the same gas deficiency, regardless of star formation properties. Furthermore, the gas-rich ones have smaller bar axial ratios or thinner bars than the gas-poor ones.
This likely suggests that the bars in gas-rich galaxies grow longer but maintain similar axial ratios over time, while they can grow fatter over time in gas-poor cases. This agrees with the assumption that the bar formation and evolution are strongly influenced by the gas fraction \citep{Cheung2013}. For example, the differences between bar formation times of massive red disk galaxies and of blue spiral ones are strongly influenced by gas \citep{Athanassoula2013}.
%there are two different processes of bar formation and evolution in gas-rich and gas-poor galaxies due to the complicated effects of gas and bulges. 
In addition, \citet{Berentzen2007} simulated the different bar evolution in gas-rich and gas-poor models. They found that bars grow in gas-poor galaxies but decline in gas-rich ones and are affected by processes such as star formation and stellar feedback. Given this, further study will be needed to make more specific and completed comparisons between gas, bars, star formation, and bulges in the future research.

\subsection{Caveats to our results}
Despite the fact that our sample galaxies span a wide range of stellar mass and the short bars below the image resolution are removed, the results still contain several potential problems, especially for the low-mass and gas-rich galaxies.

First, the spatial resolution of images may influence the fraction of detected bars and accuracy of their structural parameters. As discussed in \citet{Erwin2018}, the physically small bars are easily missed in poor-resolution images. Although our samples removed the effective radius of the bar below the typical resolution of SDSS images, there is still bias on the small end of the bar size.

Second, the method for bar detection and measurement might also contribute to the bias. As analyzed in \citet{Zou2014}, different bar measurement methods could produce systematic differences in the uncertainties of bar parameters. The bars in our samples are visually identified by the GZ2 citizen science approach, and then are measured by 2D photometric decomposition on SDSS images. Because of this, deeper images and systematic and qualitative estimation would be key in the next work.

%In observational studies, long and strong bars can be easier identified in the visual inspection, and it is more difficult for the short bars which are classified as weak bars in many studies. As we found in Figure \ref{bl_fhi_Mstar}, bar remains short in the galaxies with low stellar mass and high gas fraction, and the absolute bar length at fixed stellar mass changes little compared the disk size with increasing \hi gas fraction. Thus, the relative bar length shows a decrease with increasing gas fraction. This might partly lead to the low bar fraction found in gas-rich galaxies. This result is in agreement with simulations of barred galaxies, in which the weak bars are still able to form and grow, although at a slower rate, in gas-rich galaxies \citep{Athanassoula2013}. In the observation, they are also found abundant in the low-mass galaxies \citep{Nair2010, Erwin2018}. For example, \citet{Cervantes2017} found a positive correlation between the fraction of weak bars and the gas mass fraction.  Therefore, there are likely a substantial mount of (weak) bars in gas-rich (or low-mass) galaxies, but missed in the observation due to the small size.

%%%%%%%%%%%%%%%%%%%%%%%%%%%%%%%%%%%%%%%%%%%%%%%%%%%%%%%%%%%%%%%%%
%                        SUMMARY                                %
%%%%%%%%%%%%%%%%%%%%%%%%%%%%%%%%%%%%%%%%%%%%%%%%%%%%%%%%%%%%%%%%%
\section{Summary}
\label{sec:summary}

In this paper we present a study of the correlation between bar properties and atomic gas content of galaxies with the aim of exploring the role of \hi gas on bar evolution. The galaxies in our sample are selected from \citet{Kruk2018} with bar physical parameters measured from photometric decompositions, and are cross-matched with the ALFALFA \hi survey. In addition, we also select a comparison sample with deficient \hi fraction for given stellar masses. We explore the variations of bar properties depending on \hi gas content, stellar mass, and star formation property. Our primary results are summarized as follows:

\begin{enumerate}
	\item{The absolute bar length remains short and constant ($\rm R_{e,bar} \sim$ 2kpc) in gas-rich and low-mass galaxies and becomes longer ($\rm R_{e,bar} = $ 2-6 kpc) with increasing galactic stellar mass in gas-poor galaxies. When the stellar masses are fixed, the relative bar length ($\rm R_{e,bar}/R_{e,disk}$) shows a decrease with increasing \hi gas fraction owing to the larger disks in gas-richer galaxies. This is consistent with previous studies that the bar fraction has an anticorrelation with \hi mass fraction of disk galaxies.} 
	\item{When compared with the \hi gas-deficient galaxies at fixed stellar mass, the gas-rich galaxies ($\rm def_{\him}<0$), especially the massive ones, have longer (larger absolute length) and stronger (smaller bar axial ratio) bars.} %This indicates that bars can grow steadily in the disks with a small amount of \hi gas, while they are inhibited in a significant amount of \hi gas.} 
	\item{The bar parameters present different behaviors at low and high stellar mass bins. Bars show no significant variations for their length, axial ratio, and color in the galaxies with low stellar mass and high gas fraction, while they vary widely in the galaxies with high stellar mass and low gas fraction.} 
	\item{When splitting the samples into star forming and quiescent subgroups, we find that the star forming galaxies tend to have longer bars than the quiescent ones, with no significant difference in their the disk sizes at given stellar mass and gas fraction. This confirms that long or strong bars could enhance the star formation in the central region of galaxies.}
	\item{The bar axial ratio presents significant different variation between the star formation (star forming or quiescent) and \hi deficiency (gas-rich or gas-poor) subsets. This likely suggests that the bars in gas-rich galaxies grow longer but maintain similar axial ratios over time, while they in gas-poor cases they grow fatter over time.}
\end{enumerate}

This paper provides the analysis of the impact of varying gas fraction on the bar properties. However, the formation and growth of bars are also correlated with many other diverse components, such as the disk, bulge, and halo of host galaxies. Furthermore, the samples used in this study are mainly strong bars selected based on the classification of Galaxy Zoo 2. Further study with larger samples including both weak and strong bars will be needed to fully understand the formation and evolution of bars.

%%%%%%%%%%%%%%%%%%%%%%%%%%%%%%%%%%%%%%%%%%%%%%%%%%%%%%%%%%%%%%%%%
%                        acknowledgements                       %
%%%%%%%%%%%%%%%%%%%%%%%%%%%%%%%%%%%%%%%%%%%%%%%%%%%%%%%%%%%%%%%%%
\acknowledgements
\label{sec:acknow}
We thank the anonymous referee and editor for numerous critical comments and instructive suggestions, which have significantly improved both the content and presentation of this paper.
This work was supported by the Chinese National Natural Science Foundation grands No. 12073035, 11890693, 11873053, and 11673027, 11733006, 12090040 and 12090041 and by National Key R\&D Program of China No. 2019YFA0405501.

This work uses the data of the Arecibo Legacy Fast ALFA survey, based on observations made with the Arecibo Observatory. We acknowledge the work of the entire ALFALFA collaboration team in observing, flagging, and extracting the catalog of galaxies used in this work. 

Funding for the SDSS and SDSS-II has been provided by the Alfred P. Sloan Foundation, the Participating Institutions, the National Science Foundation, the U.S. Department of Energy, the National Aeronautics and Space Administration, the Japanese Monbukagakusho, the Max Planck Society, and the Higher Education Funding Council for England. The SDSS website is http://www.sdss.org/. The SDSS MPA-JHU catalog was produced by a collaboration of researchers (currently or formerly) from the MPA and the JHU. The team is made up of Stephane Charlot, Guinevere Kauffmann and Simon White (MPA), Tim Heckman (JHU), Christy Tremonti (University of Arizona - formerly JHU), and Jarle Brinchmann (Centro de Astrof\'isica da Universidade do Porto - formerly MPA). 
%%%%%%%%%%%%%%%%%%%%%%%%%%%%%%%%%%%%%%%%%%%%%%%%%%%%%%%%%%%%%%%%%
%                         RERERENCE                             %
%%%%%%%%%%%%%%%%%%%%%%%%%%%%%%%%%%%%%%%%%%%%%%%%%%%%%%%%%%%%%%%%%
\bibliographystyle{aasjournal}
\bibliography{}

%%%%%%%%%%%%%%%%%%%%%%%%%%%%%%%%%%%%
\end{CJK*}
\end{document}